\documentclass{iau}
\usepackage{graphicx}
\usepackage{natbib}
\usepackage{hyperref}
\usepackage[dvipsnames]{xcolor}
\usepackage{wrapfig}
\usepackage{floatrow}
\usepackage{xurl}

\newcommand{\swift}{Swift~J1818.0$-$1607}
\newcommand{\multilinecomment}[1]{}
\newcommand{\msun}{\ifmmode\mbox{M}_{\odot}\else$\mbox{M}_{\odot}$\fi}
\newcommand{\arcdeg}{\mbox{$^\circ$}}
\newcommand{\degr}{\arcdeg}
\newcommand{\fs}{\mbox{$.\!\!^{\mathrm s}$}}

\newcommand{\farcs}{\mbox{$.\!\!^{\prime\prime}$}}

\newcommand{\maspy}{$\rm mas~yr^{-1}$}
\newcommand{\kmps}{$\rm km~s^{-1}$}

\title[VLBA astrometry of \swift] 
{Probing magnetar formation channels with high-precision astrometry: The progress of VLBA astrometry of the fastest-spinning magnetar \swift}

\author[H. Ding et al.]   
{Hao Ding$^{1,2}$, Adam Deller$^{1,2}$, Marcus Lower$^{2,3}$ \and Ryan Shannon$^{1,2}$
}

\affiliation{$^1$Centre for Astrophysics and Supercomputing, Swinburne University of Technology,\\ John St., Hawthorn, VIC 3122, Australia  \\[\affilskip]
$^2$ARC Centre of Excellence for Gravitational Wave Discovery (OzGrav), Australia \\[\affilskip]
$^3$CSIRO, Space and Astronomy, Epping, NSW 1710, Australia
}

\pubyear{2022}
\volume{363}  
\pagerange{119--126}
\jname{Neutron Star Astrophysics at the Crossroads: Magnetars and the Multimessenger Revolution}
\editors{A.C. Editor, B.D. Editor \& C.E. Editor, eds.}
\begin{document}

\maketitle

\begin{abstract}
Boasting supreme magnetic strengths, magnetars are among the prime candidates to generate fast radio bursts. 
Several theories have been proposed for the formation mechanism of magnetars, but have not yet been fully tested. 
As different magnetar formation theories expect distinct magnetar space velocity distributions, high-precision astrometry of Galactic magnetars can serve as a probe for the formation theories. In addition, magnetar astrometry can refine the understanding of the distribution of Galactic magnetars. This distribution can be compared against fast radio bursts (FRBs) localized in spiral galaxies, in order to test the link between FRBs and magnetars.
\swift\ is the hitherto fastest-spinning magnetar and the fifth discovered radio magnetar. 
In an ongoing astrometric campaign, we have observed \swift\ for one year using the Very Long Baseline Array, and have determined a precise proper motion as well as a tentative parallax for the magnetar.
\end{abstract}

\section{Introduction}
\label{sec:intro}
As the most magnetized objects in the universe, magnetars may account for at least 12\% of the neutron star population \citep{Beniamini19}. However, only roughly 30 magnetars have been identified \citep{Olausen14} in our Galaxy or in the Magellanic Clouds. This discrepancy can be explained by short-lived energetic electromagnetic activities of magnetars. 

Magnetars sit at the intersection of multiple research topics. Their postulated link to fast radio bursts (FRBs) has been strengthened by the FRB-like bursts recently observed from a Galactic magnetar \citep{Andersen20,Bochenek20}. But it remains unclear whether all FRBs are originated from magnetars.
Magnetars are also strongly connected to $\gamma$-ray bursts (GRBs) through the detection of giant magnetar flares from nearby galaxies (e.g. the one in NGC~253, \citealp{Roberts21,Svinkin21}), and newborn magnetars from double neutron star mergers (e.g. \citealp{Sarin21}).
On the other hand, the formation mechanism of magnetars is yet not well understood. A few distinct magnetar formation theories have been proposed, including normal core-collapse supernovae (CCSN) of magnetic massive stars \citep{Schneider19}, accretion-induced collapse (AIC) of white dwarfs \cite{Duncan92} and double neutron star mergers \citep{Giacomazzo13,Xue19}. Most magnetar formation channels require magnetars to be born with millisecond spin periods. The only exception is the normal CCSN formation channel, where a new-born magnetar simply inherits the magnetic fields from its magnetic progenitor star \citep{Schneider19}. 

While almost all magnetars are identified with their soft $\gamma$-ray and X-ray activities, $\approx40$\% of them are also visible at optical/infrared or radio frequencies \citep{Olausen14}. This visibility allows precise astrometry for the magnetars. Historically, 7 magnetars have been precisely measured astrometrically, including 4 infrared-bright magnetars \citep{Tendulkar13} and 3 radio magnetars \citep{Deller12a,Bower15,Ding20c}.
The primary motivations of previous astrometric campaigns of magnetars are to {\bf 1)} establish supernova remnant associations and {\bf 2)} test whether magnetars receive extraordinary kick velocities ($\gtrsim1000$\,\kmps) as predicted by \citet{Duncan92}. With growing numbers of Galactic magnetars precisely measured astrometrically, new research opportunities start to emerge.

\begin{figure}
\floatbox[{\capbeside\thisfloatsetup{capbesideposition={right,center},capbesidewidth=3.5cm}}]{figure}[\FBwidth]
{\caption{Galactic latitudes of Galactic magnetars calculated from their equatorial positions provided by \citet{Olausen14}, where \swift\ is highlighted in red.
}\label{fig:gal_lat}}
{\includegraphics[width=4.4in]{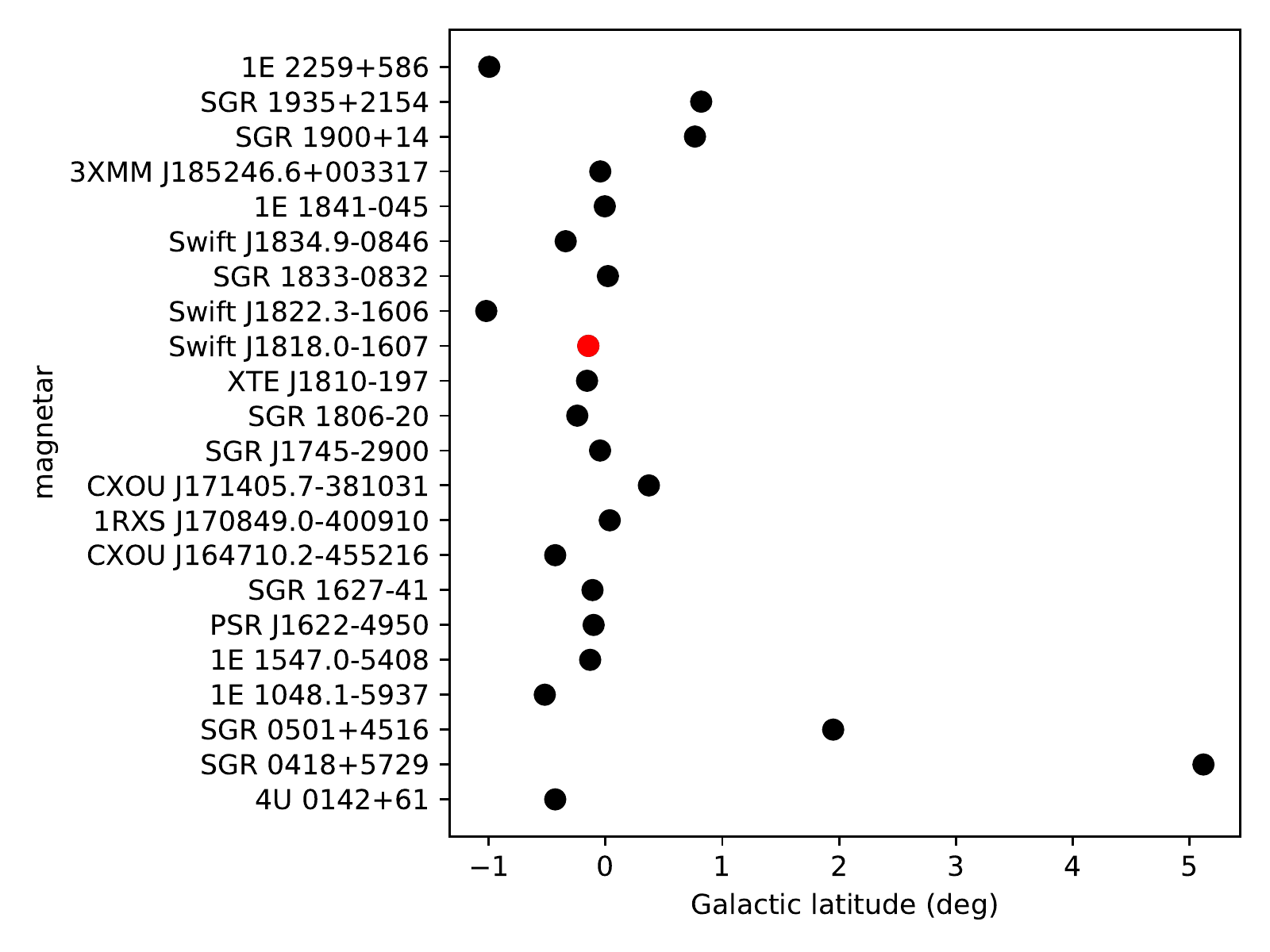}}
\end{figure}

The small sample studied by \citet{Tendulkar13}, entailing 6 astrometrically measured magnetars, offers no indication that magnetar space velocities (magnetar velocity with respect to the neighbourhood of the magnetar) follow a distribution different from that of normal pulsars \citep{Hobbs05}. This rough consistency may imply that most magnetars in spiral galaxies are born in normal CCSN similar to the ones that create typical neutron stars. The CCSN origin of magnetars is also supported by few SNR associations of magnetars (e.g. \citealp{Borkowski17}, GCN circular 16533). On the other hand, the DNS merger origin is not favored by the Galactic magnetar sample, as all Galactic magnetars are found near the Galactic plane (see Figure~\ref{fig:gal_lat}). 
For other formation channels (e.g., the AIC channel), it remains an open question if they contribute to the formation of the Galactic magnetar population, or more generally magnetars in spiral galaxies.
This question can be approached with a refined magnetar space velocity distribution, to be established with $\gtrsim$10 Galactic magnetars precisely measured astrometrically. As different formation channels may lead to distinct magnetar space velocity distributions (e.g., the AIC channel would probably give rise to relatively small magnetar space velocities), the averaged magnetar space velocity distribution could turn out to be bi-modal or even multi-modal (if there are more than one formation channel of magnetars).

Astrometry of Galactic magnetars would also play a role in the FRB study. On one hand, FRBs have been localized to specific environments (i.e., spiral arms) of spiral galaxies \citep{Mannings21}. 
On the other hand, magnetar astrometry could potentially pinpoint the 3-D magnetar location \citep{Ding20c} in the Galaxy. 
Hence, comparing the Galactic magnetar distribution against FRB sites (localized to spiral galaxies) can test the link between FRBs and magnetars.
However, Galactic magnetars precisely measured astrometrically are usually limited to the vicinity of the Solar system. 
To better infer magnetar distributions in spiral galaxies, one needs the knowledge of contributing formation channels of magnetars in the spiral galaxies (which, again, can be approached by magnetar astrometry). This is because magnetars formed in one channel are expected to follow a specific spatial distribution (e.g., magnetars born via the CCSN and DNS-merger channel are expected to be associated with, respectively, star-forming regions and stellar mass of a galaxy). 

\section{The progress of astrometry of \swift}
\label{sec:astrometry_progress}
\swift\ is the fifth discovered radio-bright \citep{Karuppusamy20} magnetar (GCN circular 27373), which is also the hitherto fastest-spinning magnetar with a spin period of 1.4\,s \citep{Enoto20}. 
Its short spin period and high spin-down rate correspond to a characteristic age of around 500\,yr \citep{champion20a}, implying its great youth.
Right after the radio detection of \swift, we launched an astrometric campaign of the magnetar using the Very Long Baseline Array (VLBA). The first VLBA observation was made at 1.6\,GHz on 20 April 2020, which did not lead to a detection \citep{Ding20b}. In response to the spectral flattening (of \swift) first noticed in July 2020 \citep[e.g.][]{Majid20}, we raised the observing frequency to 8.8\,GHz, and managed to detect \swift\ with VLBA on 19 August 2020 \citep{Ding20b}.

\begin{figure}
\floatbox[{\capbeside\thisfloatsetup{capbesideposition={right,center},capbesidewidth=2.6cm}}]{figure}[\FBwidth]
{\caption{Sky positions of \swift\ relative to the reference position $18^{\rm h}18^{\rm m}00\fs 19327$, $-16\degr07'53\farcs0095$. The positions are labeled with the observing dates in MJD. We note that the positional uncertainties presented here only reflect random errors caused by image noises. The blue curve represents the best-fit astrometric model.}\label{fig:ra_dec_model}}
{\includegraphics[width=4.4in]{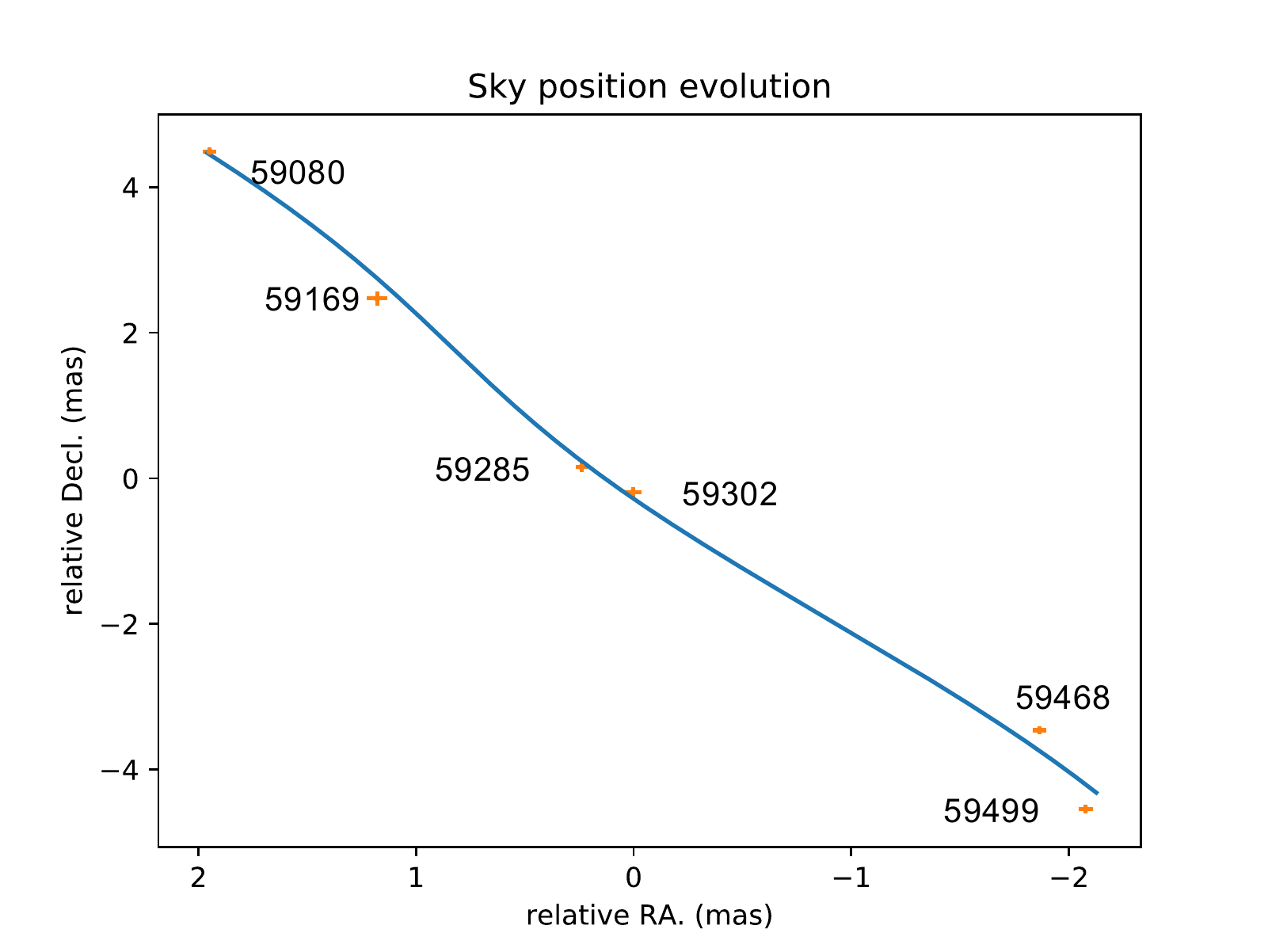}}
\end{figure}

At the time of writing, more than a year has passed since the first VLBA detection. During this period, 5 more VLBA observations have been made, all resulting in detections at sub-mas positional precision. 
We applied pulsar gating to improve image S/N, where the pulse ephemerides were generated from ongoing Parkes monitoring of \swift\ \citep{Lower20c}.
To enhance astrometric accuracy of our observations, we have employed the 1-D interpolation tactic \citep[e.g.][]{Fomalont03,Ding20c} for all the 6 VLBA observations.
After data reduction and analysis, we obtained a compelling proper motion $\mu_{\alpha}=-3.54\pm0.05$\,\maspy, $\mu_{\delta}=-7.65\pm0.09$\,\maspy, alongside a tentative (5\,$\sigma$) parallax (see Figure~\ref{fig:ra_dec_model}). Here, in order to roughly account for the systematic errors caused by atmospheric propagation effects, the uncertainty of the proper motion (as well as parallax) has been scaled by $1/\sqrt{\chi^2_{\nu}}$, where $\chi^2_{\nu}$ stands for reduced chi-square of direct parallax fitting. The final astrometric results with thorough error estimation will be obtained and discussed in a future publication, following the completion of the whole astrometric campaign that spans at least 2 years.

\section{Future prospects}
\label{sec:future prospects}
For \swift, with 6 more VLBA observations to be made in the following year, the parallax measurement is likely to become substantial ($>7\,\sigma$), which would simplify the conversion of parallax to distance (without needing to take into account the Lutz-Kelker effect, \citealp{Lutz73}). 
With regard to the whole magnetar category, to establish the magnetar space velocity distribution is mainly limited by the small sample of ($\approx11$, \citealp{Olausen14}) radio or optical/infrared magnetars. 
To expand this sample and hence accelerate the establishment of the magnetar space velocity distribution, new candidates of radio or optical/infrared magnetars, such as the newly discovered ultra-long-period sources \citep[e.g.][]{Hurley-Walker22}, are desired.

\bibliographystyle{aasjournal}
\bibliography{haoding}

\begin{thebibliography}{}
\expandafter\ifx\csname natexlab\endcsname\relax\def\natexlab#1{#1}\fi
\providecommand{\url}[1]{\href{#1}{#1}}

\bibitem[{Andersen {et~al.}(2020)Andersen, Bandura, Bhardwaj, Bij, Boyce,
  Boyle, Brar, Cassanelli, Chawla, Chen, {et~al.}}]{Andersen20}
Andersen, B., Bandura, K., Bhardwaj, M., {et~al.} 2020, arXiv preprint
  arXiv:2005.10324

\bibitem[{Beniamini {et~al.}(2019)Beniamini, Hotokezaka, van~der Horst, \&
  Kouveliotou}]{Beniamini19}
Beniamini, P., Hotokezaka, K., van~der Horst, A., \& Kouveliotou, C. 2019,
  \mnras, 487, 1426

\bibitem[{Bochenek {et~al.}(2020)Bochenek, Ravi, Belov, Hallinan, Kocz,
  Kulkarni, \& McKenna}]{Bochenek20}
Bochenek, C.~D., Ravi, V., Belov, K.~V., {et~al.} 2020, Nature, 587, 59

\bibitem[{Borkowski \& Reynolds(2017)}]{Borkowski17}
Borkowski, K.~J., \& Reynolds, S.~P. 2017, \apj, 846, 13

\bibitem[{Bower {et~al.}(2015)Bower, Deller, Demorest, Brunthaler, Falcke,
  Moscibrodzka, O'Leary, Eatough, Kramer, Lee, {et~al.}}]{Bower15}
Bower, G.~C., Deller, A., Demorest, P., {et~al.} 2015, \apj, 798, 120

\bibitem[{Champion {et~al.}(2020)Champion, Cognard, Cruces, Desvignes,
  Jankowski, Karuppusamy, Keith, Kouveliotou, Kramer, Liu,
  {et~al.}}]{champion20a}
Champion, D., Cognard, I., Cruces, M., {et~al.} 2020, Monthly Notices of the
  Royal Astronomical Society, 498, 6044

\bibitem[{Deller {et~al.}(2012)Deller, Camilo, Reynolds, \&
  Halpern}]{Deller12a}
Deller, A., Camilo, F., Reynolds, J., \& Halpern, J. 2012, The Astrophysical
  Journal Letters, 748, L1

\bibitem[{Ding {et~al.}(2020{\natexlab{a}})Ding, Deller, Lower, \&
  Shannon}]{Ding20b}
Ding, H., Deller, A.~T., Lower, M.~E., \& Shannon, R.~M. 2020{\natexlab{a}},
  ATel, 14005, 1

\bibitem[{Ding {et~al.}(2020{\natexlab{b}})Ding, Deller, Lower, Flynn,
  Chatterjee, Brisken, Hurley-Walker, Camilo, Sarkissian, \& Gupta}]{Ding20c}
Ding, H., Deller, A.~T., Lower, M.~E., {et~al.} 2020{\natexlab{b}}, \mnras,
  498, 3736.
\newblock \url{https://doi.org/10.1093/mnras/staa2531}

\bibitem[{Duncan \& Thompson(1992)}]{Duncan92}
Duncan, R.~C., \& Thompson, C. 1992, \apjl, 392, L9

\bibitem[{Enoto {et~al.}(2020)Enoto, Sakamoto, Younes, Hu, Ho, Gendreau,
  Arzoumanian, Guver, Guillot, Altamirano, {et~al.}}]{Enoto20}
Enoto, T., Sakamoto, T., Younes, G., {et~al.} 2020, ATel, 13551, 1

\bibitem[{Fomalont \& Kopeikin(2003)}]{Fomalont03}
Fomalont, E.~B., \& Kopeikin, S.~M. 2003, \apj, 598, 704

\bibitem[{Giacomazzo \& Perna(2013)}]{Giacomazzo13}
Giacomazzo, B., \& Perna, R. 2013, \apjl, 771, L26

\bibitem[{Hobbs {et~al.}(2005)Hobbs, Lorimer, Lyne, \& Kramer}]{Hobbs05}
Hobbs, G., Lorimer, D., Lyne, A., \& Kramer, M. 2005, \mnras, 360, 974

\bibitem[{Hurley-Walker {et~al.}(2022)Hurley-Walker, Zhang, Bahramian,
  McSweeney, O'Doherty, Hancock, Morgan, Anderson, Heald, \&
  Galvin}]{Hurley-Walker22}
Hurley-Walker, N., Zhang, X., Bahramian, A., {et~al.} 2022, Nature, 601, 526.
\newblock \url{https://doi.org/10.1038/s41586-021-04272-x}

\bibitem[{Karuppusamy {et~al.}(2020)Karuppusamy, Desvignes, Kramer, Porayko,
  Champion, Torne, Stappers, van~der Horst, Kouveliotou, \&
  O'Connor}]{Karuppusamy20}
Karuppusamy, R., Desvignes, G., Kramer, M., {et~al.} 2020, ATel, 13553, 1

\bibitem[{Lower {et~al.}(2020)Lower, Johnston, Shannon, Bailes, \&
  Camilo}]{Lower20c}
Lower, M.~E., Johnston, S., Shannon, R.~M., Bailes, M., \& Camilo, F. 2020,
  \mnras

\bibitem[{{Lutz} \& {Kelker}(1973)}]{Lutz73}
{Lutz}, T.~E., \& {Kelker}, D.~H. 1973, \pasp, 85, 573

\bibitem[{Majid {et~al.}(2020)Majid, Pearlman, Prince, Naudet, \&
  Bansal}]{Majid20}
Majid, W.~A., Pearlman, A.~B., Prince, T.~A., Naudet, C.~J., \& Bansal, K.
  2020, ATel, 13898, 1

\bibitem[{Mannings {et~al.}(2021)Mannings, Fong, Simha, Prochaska, Rafelski,
  Kilpatrick, Tejos, Heintz, Bannister, Bhandari, {et~al.}}]{Mannings21}
Mannings, A.~G., Fong, W.-f., Simha, S., {et~al.} 2021, \apj, 917, 75

\bibitem[{{Olausen} \& {Kaspi}(2014)}]{Olausen14}
{Olausen}, S.~A., \& {Kaspi}, V.~M. 2014, \apjs, 212, 6

\bibitem[{Roberts {et~al.}(2021)Roberts, Veres, Baring, Briggs, Kouveliotou,
  Bissaldi, Younes, Chastain, DeLaunay, Huppenkothen, {et~al.}}]{Roberts21}
Roberts, O., Veres, P., Baring, M., {et~al.} 2021, Nature, 589, 207

\bibitem[{Sarin \& Lasky(2021)}]{Sarin21}
Sarin, N., \& Lasky, P.~D. 2021, General Relativity and Gravitation, 53, 1

\bibitem[{Schneider {et~al.}(2019)Schneider, Ohlmann, Podsiadlowski, R{\"o}pke,
  Balbus, Pakmor, \& Springel}]{Schneider19}
Schneider, F.~R., Ohlmann, S.~T., Podsiadlowski, P., {et~al.} 2019, Nature,
  574, 211

\bibitem[{Svinkin {et~al.}(2021)Svinkin, Frederiks, Hurley, Aptekar,
  Golenetskii, Lysenko, Ridnaia, Tsvetkova, Ulanov, Cline,
  {et~al.}}]{Svinkin21}
Svinkin, D., Frederiks, D., Hurley, K., {et~al.} 2021, Nature, 589, 211

\bibitem[{Tendulkar {et~al.}(2013)Tendulkar, Cameron, \&
  Kulkarni}]{Tendulkar13}
Tendulkar, S.~P., Cameron, P.~B., \& Kulkarni, S.~R. 2013, \apj, 772, 31

\bibitem[{Xue {et~al.}(2019)Xue, Zheng, Li, Brandt, Zhang, Luo, Zhang, Bauer,
  Sun, Lehmer, {et~al.}}]{Xue19}
Xue, Y., Zheng, X., Li, Y., {et~al.} 2019, Nature, 568, 198

\end{thebibliography}
\setlength{\itemsep}{0mm}
\small

\end{document}